\documentclass[12pt]{article}
 \usepackage[cp1251]{inputenc} %- для WiN
 \usepackage[english]{babel}
 \usepackage{amsmath, latexsym, amssymb, bm, array, graphics, eucal,
 amsfonts,}%mathscr
 \makeatletter
 \renewcommand{\@biblabel}[1]{#1.\hfill}

 \renewcommand{\Re}{\mathop{\rm Re}}
 \renewcommand{\Im}{\mathop{\rm Im\,}}

\newcommand{\mc}[1]{\mathcal{#1}}
\newcommand{\E}{\mc{E}}
 \makeatother
 
\usepackage[dvips]{graphicx}
\usepackage[flushleft,small]{caption2}

\begin{document}
 \thispagestyle{empty}
 \renewcommand{\abstractname}{\ }
 \large
 \renewcommand{\refname}{\begin{center} REFERENCES\end{center}}
\newcommand{\const}{\mathop{\rm const\, }}
 \begin{center}
\bf TRANSVERSE ELECTRIC CONDUCTIVITY OF QUANTUM COLLISIONAL
MAXWELLIAN PLASMA
\end{center}\medskip
\begin{center}
  \bf  A. V. Latyshev\footnote{$avlatyshev@mail.ru$},
  A. A. Yushkanov\footnote{$yushkanov@inbox.ru$}
\end{center}\medskip

\begin{center}
{Faculty of Physics and Mathematics,\\ Moscow State Regional
University,  105005,\\ Moscow, Radio str., 10--A}
\end{center}\medskip

\begin{abstract}
Formulas for calculation of transverse dielectric function and
transverse electric conductivity in quantum collisional Maxwellian
plasma are obtained.
The Wigner -- Vlasov -- Boltzmann kinetic equation with collision
integral in BGK (Bhatnagar, Gross and Krook) form
in coordinate space is used. Various special cases are
in\-ves\-ti\-gated.
Comparison with Lindhard's formula has been carried out. \medskip

{\bf Key words:} collisional plazma,
electric conductivity, dielectric permeability, Lindhard
formula.\medskip

PACS numbers:  52.25.Dg Plasma kinetic equations,
52.25.-b Plasma properties

\end{abstract}

\begin{center}\bf
  1. Introduction
\end{center}

In the present work formulas for calculation of electric
conductivity and dielectric permeability in quantum collisional
Maxwellian plasma  are deduced.

During the derivation of the kinetic equation we generalize the
approach, developed by Klimontovich and Silin \cite {Klim}.

Dielectric permeability in the collisionless quantum gaseous plasma
was studied by many authors (see, for example, \cite {Klim} --
\cite{Shukla2}).

In the work \cite {Manf}, where the one-dimensional case
of the quantum plasma is investigated, the importance of derivation
of dielectric permeability with use of the quantum kinetic equation
with collision integral in the form of BGK -- model (Bhatnagar,
Gross, Krook) \cite{BGK}, \cite{Opher} was noted.

The present work is devoted to the performance of this task.

A dielectric permeability is one of the most significant
characteristics of a plasma. This quantity is necessary for
description of the skin effect \cite{Gelder},
for analysis of surface plasmons \cite{Fuchs},
for description of the process of propagation and damping of the
transverse plasma oscillations \cite{Shukla2},
the mechanism of electromagnetic
waves penetration in plasma \cite{Shukla1}, and
for analysis of other problems of plasma physics \cite{Fuchs2},
\cite{Dressel}, \cite{Wier}, \cite{Brod} and \cite{Manf2}.

In the present work the Wigner --- Vlasov --- Boltzmann kinetic
equation  with collision integral in the form of relaxation
$\tau$ -- model in coordinate space is used.

This equation is considered about  the Wigner function,
which is quantum analog of classical distribution function.
The Wigner function  has been entered in work \cite {Wigner}
and then it was investigated in works \cite {Tatarskii} -- \cite {Kozlov}.

Kliewer and Fuchs were the first who have noticed \cite {Kliewer},
that the dielectric function for quantum plasma deduced by Lindhard
in collisional case does not pass into dielectric function for
classical plasma in the limit when Planck's constant $\hbar$
tends to zero. This means, that dielectric Lindhard's function
does not take into account electron collisions correctly. Kliewer
and Fuchs have corrected Lindhard's dielectric function "by hands" \,
so that it passed into classical one under condition $ \hbar \to 0$.

In the works  \cite{Fuchs}, \cite{Fuchs2} the dielectric function
received by them (Kliewer and Fuchs)
was applied to consideration of various questions
of  metal optics.

In the work \cite{Mermin} the correct account of collisions in
framework of the relaxation model in electron momentum space
for the case of longi\-tu\-di\-nal dielectric function has been
carried out. At
the same time the correct account of influence of collisions for
transverse dielectric function has not been implemented till
now.

The aim of the present work is the elimination of this lacuna.

\begin{center}
  \bf 2. Transverse conductivity in maxwellian plasma
\end{center}

In a considered case of plasma locally equilibrium maxwellian
distribution function looks like (see \cite{Lat1}):
$$
f_{\rm eq} \equiv f^{(0)}=\dfrac{N_0}{v_T \pi^{3/2}}
\exp\left\{-\Big(\mathbf{P}-\dfrac{e}{cp_T}\mathbf{A}
(\mathbf{r},t)\Big)^2-\dfrac{eU}{k_BT}\right\}.
\eqno{(1.1)}
$$

Here $v_T=\sqrt{\dfrac{2k_BT}{m}}$ is the thermal electron velocity,
$m$ is the electron mass, $e$ is the electron charge,
$p_T=mv_T$ is the thermal electron momentum,
$\mathbf{P}$ is the dimensinless electron momentum,
$N_0$ is the electron number density in equilibrium state,
$\mathbf{A}(\mathbf{r},t)$ is the vector potential, $U=U(\mathbf{r},t)$
is the scalar potential (which is considered further equal  to zero),
$c$ is the light velocity, $k_B$ is the Boltzmann constant, $T$
is the plasma temperature,
$f_0(P)$ is the absolute Maxwell distribution,
$$
f_0(P)=\dfrac{N_0}{v_T^3 \pi^{3/2}}e^{-P^2}.
$$

The Wigner --- Vlasov functional in this case equals to \cite{Lat1}:
$$
W[f]=\big(\mathbf{P\,A}\big)\dfrac{iev_T}{c\hbar}(f_0^+(\mathbf{P})-
f_0^-(\mathbf{P})),
\eqno{(1.2)}
$$
where
$$
f_0^{\pm}\equiv f_0^{\pm}(\mathbf{P})=\dfrac{N_0}{v_T^3 \pi^{3/2}}
\exp\Big\{-\Big(\mathbf{P}\mp
\dfrac{\mathbf{q}}{2}\Big)^2\Big\},
$$
$\mathbf{q}$ is the dimensionless wave vector, $k_T$ is the wave
number corresponding to thermal movement of molecules,
$$
\mathbf{q}=\dfrac{\mathbf{k}}{k_T}, \qquad
k_T=\dfrac{p_T}{\hbar}=\dfrac{mv_T}{\hbar}.
$$

We consider the kinetic equation with the Wigner --- Vlasov
functional (1.2)
$$
\dfrac{\partial f}{\partial t}+\mathbf{v}\dfrac{\partial f}{\partial
\mathbf{r}}=\nu (f_{\rm eq}-f)+(\mathbf{P A})\dfrac{iev_T}{c\hbar}
(f_0^+-f_0^-).
\eqno{(1.3)}
$$

Here $\nu$ is the effective electron collision  frequency.

Linearization of equilibrium distribution function (1.1) gives:
$$
f_{\rm eq}=f_0(P)+f_0(P)\dfrac{2e}{cp_T}(\mathbf{P A}).
$$

Therefore we seek the solution of equation (1.3) in the form
$$
f=f_0(P)+f_0(P)\dfrac{2e}{cp_T}(\mathbf{P A})+f_0(P)\dfrac{2e}{cp_T}
(\mathbf{PA})h(\mathbf{P}).
\eqno{(1.4)}
$$

Substituting (1.4) in (1.3) we receive
$$
(\mathbf{PA})f_0(P)\dfrac{2e}{cp_T}(\nu-i\omega+i\mathbf{kv})=
\dfrac{2ie}{cp_T}(\mathbf{PA})(\omega-\mathbf{kv})f_0(P)
$$
$$
+(\mathbf{PA})\dfrac{iev_T}{c\hbar}(f_0^+-f_0^-).
\eqno{(1.5)}
$$

Let's replace in the equation (1.5) $\mathbf{v}$ on
$$
\mathbf{v}=\dfrac{\mathbf{p}}{m}-\dfrac{e\mathbf{A}}{mc}=
v_T\mathbf{P}-\dfrac{e\mathbf{A}}{mc}
$$
and in the received equation we will produce linearization on the vector
field $\mathbf{A}$. We receive the equation from which it is found
$$
(\mathbf{PA})h(\mathbf{P})f_0(P)\dfrac{2ie}{cp_T}=(\mathbf{PA})
\dfrac{\dfrac{2e}{cp_T}(\omega-v_T\mathbf{kP})+
\dfrac{iev_T}{c\hbar}(f_0^+-f_0^-)}{\nu-i\omega+iv_T\mathbf{kP}}
\eqno{(1.6)}
$$

Substituting (1.6) in (1.4) we receive
$$
f=f_{\rm eq}+\dfrac{2ie}{cp_T}(\mathbf{PA})
\dfrac{(\omega-v_T\mathbf{kP})f_0(P)+\dfrac{\E_T}{\hbar}(f_0^+-f_0^-)}
{\nu-i\omega+iv_T\mathbf{kP}},
\eqno{(1.7)}
$$
where $\E_T$ is the thermal electron energy,
$$
\E_T=\dfrac{mv_T^2}{2}.
$$

%Take into account, that the
Electric current in an equilibrium condition
is equal to zero \cite{Lat1}. Then we receive using
(1.7)
$$
\mathbf{j}(\mathbf{r},t)=e\int \mathbf{v}f(\mathbf{r}, \mathbf{v},t)d^3v
$$
$$
=\dfrac{2ie^2}{cp_T}\int \mathbf{v}(\mathbf{PA})
\dfrac{(\omega-v_T\mathbf{kP})f_0(P)+(\E_T/\hbar)(f_0^+-f_0^-)}
{\nu-i\omega+iv_T\mathbf{kP}}d^3v.
\eqno{(1.8)}
$$

After linearization on the vector field we obtain
$$
\mathbf{j}=\dfrac{2ie^2v_T^4}{cp_T}\int \mathbf{P}(\mathbf{PA})
\dfrac{(\omega-v_T\mathbf{kP})f_0(P)+(\E_T/\hbar)(f_0^+-f_0^-)}
{\nu-i\omega+iv_T\mathbf{kP}}d^3P.
\eqno{(1.9)}
$$

We take unit vector $\mathbf{e}_1=\dfrac{\mathbf{A}}{A}$,
directed alongwise to the vector $\mathbf{A}$. Then equality (1.9)
may be rewritten in the form
$$
\mathbf{j}(\mathbf{r},t)=\dfrac{2ie^2v_T^3 A(\mathbf{r},t)}
{cm}
\int \mathbf{P}\big(\mathbf{P}\mathbf{e}_1\big)S(P^2,
\mathbf{k}_1\mathbf{P})\,d^3P,
\eqno{(1.9')}
$$
where $\mathbf{k}_1=v_T\tau \mathbf{k}=l \mathbf{k}$ is the
dimensionless wave vector,
$$
S(P^2,\mathbf{k}_1\mathbf{P})=
\dfrac{(\omega\tau-\mathbf{k}_1\mathbf{P})f_0(P)+(\E_T/\hbar)(f_0^+-f_0^-)}
{1-i\omega\tau+i\mathbf{k}_1\mathbf{P}}.
$$

We take other unit vector
$\mathbf{e}_2$, which is perpendicular to the vector $\mathbf{k}_1$, i.e.
$$\qquad \qquad \mathbf{e}_2=\dfrac{\mathbf{A} \times \mathbf{k}_1}
{|\mathbf{A} \times \mathbf{k}_1|}=
\dfrac{\mathbf{A} \times \mathbf{k}_1}{Ak_1},
$$
where $\mathbf{A} \times \mathbf{k}_1$ is the vector product.

We expand the vector $\mathbf{P}$ on three orthogonal directions
$\mathbf{e}_1$, $\mathbf{e}_2$ and
$\mathbf{n}=\dfrac{\mathbf{k}_1}{k_1}=\dfrac{\mathbf{k}}{k}$:
$$
\mathbf{P}=(\mathbf{Pn})\mathbf{n}+(\mathbf{P}\mathbf{e}_1)
\mathbf{e}_1+(\mathbf{P}\mathbf{e}_2)\mathbf{e}_2.
$$

We receive with the help of this expansion
$$
(\mathbf{PA})\mathbf{P}=A(\mathbf{P}\mathbf{e}_1)\mathbf{P}=
$$
$$
=A(\mathbf{P}\mathbf{e}_1)(\mathbf{Pn})\mathbf{n}+
A(\mathbf{P}\mathbf{e}_1)^2\mathbf{e}_1+A(\mathbf{P}\mathbf{e}_1)
(\mathbf{P}\mathbf{e}_2)\mathbf{e}_2.
$$

Substituting this decomposition in $(1.9')$, and, considering,
that integrals from odd functions on a symmetric interval are equal
to zero, we receive:
$$
\mathbf{j}(\mathbf{r},t)=\dfrac{2ie^2v_T^3 \mathbf{A}(\mathbf{r},t)}
{cm}\int (\mathbf{P}\mathbf{e}_1)^2
S(P^2,\mathbf{k}_1\mathbf{P})\,d^3P.%dP_ndP_{e_1}dP_{e_2}.
\eqno{(1.10)}
$$

In view of symmetry value of integral will not change,
if the vector $\mathbf{e}_1$ to replace with any other unit
vector $\mathbf{e}_2$, perpendicular to the vector $\mathbf{k}_1$.
Therefore
$$
\int \big(\mathbf{P}\mathbf{e}_1\big)^2S(P^2,\mathbf{k}_1\mathbf{P})d^3P=
\int \big(\mathbf{P}\mathbf{e}_2\big)^2S(P^2,\mathbf{k}_1\mathbf{P})d^3P
$$
$$
=\dfrac{1}{2}\int \Big[\Big( \mathbf{P}\mathbf{e}_1\Big)^2
+\Big(\mathbf{P}\mathbf{e}_2\Big)^2\Big]
S(P^2,\mathbf{k}_1\mathbf{P})d^3P.
$$

Hence, for current density we have:
$$
\mathbf{j}(\mathbf{r},t)=
\dfrac{ie^2v_T^3 \mathbf{A}(\mathbf{r},t)}{cm}\int \Big[(\mathbf{Pe}_1)^2+
(\mathbf{Pe}_2)^2\Big]S(P^2,\mathbf{k}_1\mathbf{P})d^3P.
$$

Let's notice, that the value
$P^2=\mathbf{P}\mathbf{P}$ is equal to:
$$
P^2=(\mathbf{P}\mathbf{e}_1)^2+(\mathbf{P}\mathbf{e}_2)^2+
(\mathbf{Pn})^2,
$$
whence
$$
\big(\mathbf{P}\mathbf{e}_1\big)^2
+\big(\mathbf{P}\mathbf{e}_2\big)^2=P^2-
\dfrac{(\mathbf{P}\mathbf{k}_1)^2}{k_1^2}=P^2-(\mathbf{Pn})^2=
P_\perp^2,
$$
where $P_\perp$ is the projection of vector $\mathbf{P}$ on a
straight line, perpendicular to plane $(\mathbf{e}_1, \mathbf{e}_2)$.

Then for current density we receive the following expression
$$
\mathbf{j}(\mathbf{r},t)=
\dfrac{ie^2v_T^3 \mathbf{A}(\mathbf{r},t)}{cm}\int
P_\perp^2\,S(P^2,\mathbf{k}_1\mathbf{P})d^3P.
$$

We consider the connection between electric field and potentials
$$
\mathbf{E}(\mathbf{r},t)=-\dfrac{1}{c}
\dfrac{\partial \mathbf{A}(\mathbf{r},t)}{\partial t}-
\dfrac{\partial U(\mathbf{r},t)}{\partial \mathbf{r}},
$$
or
$$
\mathbf{E}(\mathbf{r},t)=\dfrac{i\omega}{c}\mathbf{A}(\mathbf{r},t).
$$

Hence, the current is connected with vector potential as:
$$
\mathbf{j}(\mathbf{r},t)=\sigma_{tr}\mathbf{E}(\mathbf{r},t)=
\sigma_{tr}\dfrac{i\omega}{c}\mathbf{A}(\mathbf{r},t).
$$

Replacing a current in the left part of this equality expression
through a field, we receive
$$
\sigma_{tr}\dfrac{i \omega}{c}\mathbf{A}(\mathbf{r},t)=
\dfrac{ie^2v_T^3 \mathbf{A}(\mathbf{r},t)}{cm}\int
P_\perp^2\,S(P^2,\mathbf{k}_1\mathbf{P})d^3P.
$$

From here we receive expression for transverce electric conductivity
$$
\sigma_{tr}=
\dfrac{e^2v_T^3}{m\omega}\int
P_\perp^2\,S(P^2,\mathbf{k}_1\mathbf{P})d^3P.
$$

Let's replace in this expression the absolute Maxwell distribution
by its  explicit  representation. We receive that:
$$
\sigma_{tr}=\dfrac{e^2N_0}{m\omega \pi^{3/2}}\int
\dfrac{(\omega-v_T\mathbf{kP})e^{-P^2}+\E_T(e^{-P^2_+}-
e^{-P^2_-})/\hbar}{\nu-i\omega+iv_T\mathbf{kP}}P_\perp^2d^3P.
\eqno{(1.11)}
$$

Let's transform the formula (1.11) %, having allocated
with the use of standard expression for static electric conductivity
$\sigma_0=(e^2N_0)/(m\nu)$:
$$
\sigma_{tr}= \dfrac{\sigma_0}{\pi^{3/2}}\int
\Bigg[\Big(1-\dfrac{\mathbf{k}_1\mathbf{P}}{\omega \tau}\Big)e^{-P^2}+
\dfrac{\E_T}{\hbar \omega}\Big(e^{-P_+^2}-e^{-P_-^2}\Big)\Bigg]
%\times
%$$
%$$\times
\dfrac{P_\perp^2\,d^3P}{1-i\omega \tau+i \mathbf{k}_1\mathbf{P}}.
\eqno{(1.12)}
$$

On the basis of formulas (1.11) or (1.12) we will write the formula
for  transverse dielectric permeability
 $\varepsilon_{tr}=
1+\dfrac{4\pi i}{\omega}\sigma_{tr}$ in Maxwellian plasma
$$
\varepsilon_{tr}=1+\dfrac{i\omega_p^2}{\omega^2 \pi^{3/2}}\int
\Bigg[\Big(\omega \tau-\mathbf{k}_1\mathbf{P}\Big)e^{-P^2}+
\dfrac{\E_T}{\hbar \nu}\Big(e^{-P_+^2}-e^{-P_-^2}\Big)\Bigg]
%\times
%$$
%$$\times
\dfrac{P_\perp^2\,d^3P}{1-i\omega \tau+i \mathbf{k}_1\mathbf{P}}.
$$

Let's present (1.11) in the form of the sum
$$
\sigma_{tr}=\sigma_{tr}^{\rm classic}+\sigma_{tr}^{\rm quant}.
\eqno{(1.13)}
$$

Here
$$
\sigma_{tr}^{\rm classic}=\sigma_0 f_{\rm classic},
\eqno{(1.14)}
$$
where
$$
f_{\rm classic}=\dfrac{1}{\pi^{3/2}}\int \dfrac{e^{-P^2}P_\perp^2d^3P}
{1-i\omega\tau+il \mathbf{kP}},
\eqno{(1.15)}
$$
and
$$
\sigma_{tr}^{\rm quant}=\sigma_0 f_{\rm quant},
\eqno{(1.16)}
$$
where
$$
f_{\rm quant}=\dfrac{1}{\pi^{3/2}}\int
\Bigg\{ -\dfrac{v_T}{\omega}\mathbf{kP}e^{-P^2}\hspace{7cm}
$$$$+\dfrac{k_Tv_T}{2\omega}
(e^{-P^2_+}-e^{-P^2_-})\Bigg\}\dfrac{P_\perp^2d^3P}{1-i\omega\tau+il
\mathbf{kP}}.
\eqno{(1.17)}
$$

We will present the formula (1.16)  in the form of the sum of two
terms
$$
\sigma^{\rm quant}=\sigma_1+\sigma_2.
\eqno{(1.18)}
$$

In the formula (1.18) following designations are entered
$$
\sigma_1=-\dfrac{l}{\omega\tau}\dfrac{1}{\pi^{3/2}}\int
\dfrac{\mathbf{kP}e^{-P^2}P_\perp^2\,d^3P}{1-i\omega\tau+il\mathbf{kP}},
\eqno{(1.19)}
$$
and
$$
\sigma_2=\dfrac{lk_T}{2\omega\tau}\dfrac{1}{\pi^{3/2}}\int
\dfrac{(e^{-P_+^2}-e^{-P_-^2})P_\perp^2\,d^3P}{1-i\omega\tau+il\mathbf{kP}}.
\eqno{(1.20)}
$$

After obvious replacement of variables we  receive
$$
\dfrac{e^{-(\mathbf{P\mp
\mathbf{q}/2})^2}}{1-i\omega\tau+il\mathbf{kP}}\to
\dfrac{e^{-P^2}}{1-i\omega\tau+il\mathbf{k}(\mathbf{P}\pm
\mathbf{q}/2)}.
$$

Let's notice, that
$$
\dfrac{1}{1-i\omega\tau+il\mathbf{kP}+il\mathbf{kq}/2}-
\dfrac{1}{1-i\omega\tau+il\mathbf{kP}-il\mathbf{kq}/2}
$$
$$
=
\dfrac{-il\mathbf{kq}}{(1-i\omega\tau+il\mathbf{kP})^2+(l\mathbf{kq}/2)^2}.
$$

By means of two last  equalities we receive expression for
$\sigma_2$:
$$
\sigma_2=-i\sigma_0\dfrac{(lk)^2}{\omega\tau}\dfrac{1}{\pi^{3/2}}
\int \dfrac{e^{-P^2}P_\perp^2\,d^3P}
{(1-i\omega\tau+il\mathbf{kP})^2+(l\mathbf{kq}/2)^2}.
\eqno{(1.21)}
$$

Let's consider, that $\mathbf{k}=k\mathbf{n},\;\mathbf{kP}=k\mathbf{nP}=
kP_n $, thus $P^2=P_n^2+P_\perp^2$, besides,
$$
\mathbf{kq}=k\mathbf{nq}=k \mathbf{n}\dfrac{\mathbf{k}}{k_T}=
k^2\dfrac{\mathbf{n^2}}{k_T}=\dfrac{k^2}{k_T}=kq.
$$

Therefore we may write down the following expression  for the component of transverse conductivity
$\sigma_2$:
$$
\sigma_2=i\sigma_0\dfrac{\nu}{\omega}\dfrac{1}{\pi^{3/2}}\int
\dfrac{e^{-P^2}P_\perp^2\,d^3P}{(P_n-z/q)^2-(q/2)^2},
\eqno{(1.22)}
$$
where dimensionless parametres are entered
$$
z=x+iy=\dfrac{\omega+ \nu}{k_Tv_T}, \qquad
x=\dfrac{\omega}{k_Tv_T}, \qquad y=\dfrac{ \nu}{k_Tv_T}.
$$

Let's notice, that the formula (1.12) can be deduced
from the general formula for transverse conductivity of quantum plasma
at arbitrary degree degeneration of electronic gas (see, for example,
\cite {Lat1})
$$
\dfrac{\sigma_{tr}}{\sigma_0}=\dfrac{1}{4\pi f_2(\alpha)}
\int \Bigg\{\Big(1-\dfrac{\mathbf{k}_1\mathbf{P}}{\omega \tau}\Big)g(P)
\hspace{5cm}
$$
$$
\hspace{2cm}+\dfrac{\E_T}{\hbar \omega}\Big(f_F^+-f_F^-\Big)\Bigg\}
\dfrac{P_\perp^2\,d^3P}{1-i\omega \tau +i
\mathbf{k}_1\mathbf{P}}.
\eqno{(1.23)}
$$

In the formula (1.23) following designations are entered
$$
f_2(\alpha)=\int\limits_{0}^{\infty}x^2f_F(x)dx=
\int\limits_{0}^{\infty}\dfrac{x^2}{1+e^{x^2-\alpha}},
$$
$$
g(P)=\dfrac{e^{P^2-\alpha}}{(1+e^{P^2-\alpha})^2},
$$
where $\alpha$ is the dimensionless chemical potential,
$\alpha=\dfrac{\mu}{k_BT}$.

Let's notice that by $\alpha\to -\infty$
$f_2(\alpha)=e^\alpha\dfrac{\sqrt{\pi}}{4}$.
Therefore we have two following  limiting transitions
$$
\lim\limits_{\alpha\to -\infty}\dfrac{g(P)}{4\pi f_2(\alpha)}=
\dfrac{e^{-P^2}}{\pi^{3/2}},
$$
$$
\lim\limits_{\alpha\to -\infty}
\dfrac{f_F^+(\mathbf{P})-f_F^-(\mathbf{P})}{4\pi f_2(\alpha)}=
\dfrac{e^{-P_+^2}-e^{-P_-^2}}{\pi^{3/2}}.
$$

Taking into account two last limiting transitions it is clear,
that the formula
(1.23) at $\alpha\to-\infty$ passes in the formula (1.12).

\begin{center}
  \bf 3. Special cases and properties of transverse conductivity
\end{center}

When wave number $k=0$ from the formula (1.12) we obtain the known
classical formula for conductivity:
$$
\sigma_{tr}(k=0)=\dfrac{\sigma_0}{\pi^{3/2}}
\int \dfrac{e^{-P^2}P_\perp^2\,d^3P}
{1-i\omega\tau}=\dfrac{\sigma_0}{1-i\omega\tau}=
\sigma_0\dfrac{\nu}{\nu-i\omega},
$$
because
$$
\dfrac{1}{\pi^{3/2}}\int e^{-P^2}(P_y^2+P_z^2)d^3P=1.
$$

We will spread out  the expression  in  braces (1.17) on degrees of
$\mathbf{q}$
$$
\dfrac{v_T}{\omega}\Big\{-\mathbf{Pk}e^{-P^2}+
\dfrac{k_T}{2}(e^{-P^2_+}-e^{-P^2_-})\Big\}
$$
$$
=\dfrac{v_T}{\omega}e^{-P^2}\dfrac{1}{6}k_T (\mathbf{P q})
[(\mathbf{P q})^2-\dfrac{3}{2}q^2]
$$$$=\dfrac{v_Tk_T}{6\omega}e^{-P^2}q^3
P_n(P_n^2-\dfrac{3}{2}), \qquad P_n=\mathbf{P n}.
$$

On the basis of this decomposition at small $q$ we receive
$$
f_{\rm quant}=\dfrac{v_Tk_T}{6\omega \pi^{3/2}}\int
 (\mathbf{P q})
[(\mathbf{P q})^2-\dfrac{3}{2}q^2]\dfrac{P_\perp^2d^3P}{1-i\omega\tau+il
\mathbf{Pk}}.
\eqno{(2.1)}
$$

By means of (2.1) from the formula (1.12) at small $q $ we obtain
$$
\dfrac{\sigma_{tr}}{\sigma_0}=\dfrac{1}{\pi^{3/2}}
\int \dfrac{e^{-P^2}P_\perp^2d^3P}{1-i\omega\tau +il \mathbf{k}
\mathbf{P}}\hspace{6cm}
$$$$+
\dfrac{k_Tv_T}{6\omega \pi^{3/2}}\int
\dfrac{(\mathbf{P q})[(\mathbf{P q})^2-\dfrac{3}{2}q^2]
P_\perp^2d^3P}{1-i\omega\tau +il \mathbf{k}
\mathbf{P}}.
\eqno{(2.2)}
$$

From the formula (2.2) follows, that at $\mathbf{q}\to0$ (or,
 when Planck's constant tends to zero $\hbar\to 0$),
the formula for transverse conductivity of quantum plasma passes in
the formula of transverse conductivity of classical plasma.

Vector $\mathbf{k}$ we will direct along an axis $x$,
$\mathbf {k}=k\{1,0,0\}$, then
$$
\mathbf{Pk}=kP_x,  \qquad
P_{\pm}^2=\Big(P_x\pm \dfrac{q}{2}\Big)^2.
$$

Hence, the formula (1.12) can be rewritten in the form
$$
\dfrac{\sigma_{tr}}{\sigma_0}=\dfrac{1}{\pi^{3/2}}\int
\Bigg[\Big(1-\dfrac{v_T}{\omega}kP_x\Big)e^{-P^2}\hspace{4cm}$$$$+
\dfrac{k_Tv_T}{2\omega}\Big(e^{-P_+^2}-e^{-P_-^2}\Big)\Bigg]
\dfrac{(P_y^2+P_z^2)d^3P}{1-i\omega\tau+ilkP_x}.
\eqno{(2.3)}
$$

The internal double integral can be easy calculated in the polar
coordinates in a plane $(P_y,P_z)$:
$$
\int e^{-P_y^2-P_z^2}(P_y^2+P_z^2)dP_ydP_z=\pi.
$$

Hence, according to (2.3) the transverse conductivity is expressed
by one-dimensional integral:
$$
\dfrac{\sigma_{tr}}{\sigma_0}=\dfrac{1}{\sqrt{\pi}}
\int\limits_{-\infty}^{\infty}
\Bigg\{\Big(1-\dfrac{v_T}{\omega}kP_x\Big)e^{-P_x^2}\hspace{6cm}$$$$+
\dfrac{k_Tv_T}{2\omega}\Big[e^{-(P_x-q/2)^2}-e^{-(P_x+q/2)^2}\Big]\Bigg\}
\dfrac{dP_x}{1-i\omega\tau+ilkP_x}.
\eqno{(2.4)}
$$

The formula (2.4) can be transformed to the form
$$
\dfrac{\sigma_{tr}}{\sigma_0}=\dfrac{1}{\sqrt{\pi}}
\int\limits_{-\infty}^{\infty}
\Big\{\Big(1-\dfrac{lk}{\omega\tau}\mu\Big)e^{-\mu^2}\hspace{5cm}$$$$+
\dfrac{lk_Te^{-q^2/4}}{2\omega\tau}\big(e^{q\mu}-e^{-q\mu}\big)\Big\}
\dfrac{e^{-\mu^2}d\mu}{1-i\omega\tau+ilk\mu}.
\eqno{(2.5)}
$$

Let's us notice that at small  $q$
$$
e^{-q^2/4}\big(e^{q\mu}-e^{-q\mu}\big)=2q\mu+\mu(\mu^2-\dfrac{3}{2})
\dfrac{q^3}{3}+\cdots .
$$

Hence, at small $q=k/k_T $ we receive
$$
\sigma_{tr}=\sigma^{\rm classic}_{tr}+
\dfrac{\sigma_0\hbar^2k^3}{6\omega m^2v_T\sqrt{\pi}}
\int\limits_{-\infty}^{\infty}
\dfrac{e^{-\mu^2}\mu(\mu^2-3/2)d\mu}{1-i\omega\tau+ilk\mu}.
\eqno{(2.6)}
$$

In the formula (2.6) the first term is the transverse conductivity
in classical plasma
$$
\sigma^{\rm classic}_{tr}=\dfrac{\sigma_0}{\sqrt{\pi}}
\int\limits_{-\infty}^{\infty}\dfrac{e^{-\mu^2}d\mu}{1-i\omega\tau+ilk\mu}.
$$

Let's transform the denominator of the formula (2.4)
$$
\dfrac{1}{1-i\omega\tau+ilk\mu}=\dfrac{1}{ilk}\dfrac{1}{\mu+
(1-i\omega\tau)/(ilk)}=\dfrac{1}{ilk}\dfrac{1}{\mu-z/q},
$$
where  dimensionless parametres are entered
$$
x=\dfrac{\omega}{k_Tv_T}, \qquad y=\dfrac{\nu}{k_Tv_T},
\qquad z=x+iy=\dfrac{\omega+i \nu}{k_Tv_T}.
$$

By means of these designations we receive
$$
\sigma_{tr}^{\rm
classic}=-i\sigma_0\dfrac{y}{q}t\Big(\dfrac{z}{q}\Big),
\eqno{(2.7)}
$$
where
$$
t(z)=\dfrac{1}{\sqrt{\pi}}\int\limits_{-\infty}^{\infty}
\dfrac{e^{-\tau^2}d\tau}{\tau-z}.
$$

Let's rewrite the formula (2.6) in dimensionless variables
$$
\dfrac{\sigma_{tr}}{\sigma_0}=-i\dfrac{y}{q}t\Big(\dfrac{z}{q}\Big)-
i\dfrac{y}{x}q^2\dfrac{1}{\sqrt{\pi}}\int\limits_{-\infty}^{\infty}
\dfrac{e^{-\tau^2}\tau(\tau^2-3/2)d\tau}{\tau-z/q}.
\eqno{(2.8)}
$$

Let's transform the formula (2.8) in the form
$$
\dfrac{\sigma_{tr}}{\sigma_0}=-i\dfrac{y}{q}t\Big(\dfrac{z}{q}\Big)-
i\dfrac{y}{x}q^2\Big[\dfrac{1}{2}+\Big(\dfrac{z^2}{q^2}-\dfrac{3}{2}\Big)
\lambda_C\Big(\dfrac{z}{q}\Big)\Big].
\eqno{(2.9)}
$$

In (2.9) $\lambda_C(z)$ is the plasma dispersion function
entered by Van Campen
$$
\lambda_C(z)=1+zt(z)=1+\dfrac{z}{\sqrt{\pi}}\int\limits_{-\infty}^{\infty}
\dfrac{e^{-\tau^2}d\tau}{\tau-z}=
\dfrac{1}{\sqrt{\pi}}\int\limits_{-\infty}^{\infty}
\dfrac{\tau e^{-\tau^2}d\tau}{\tau-z}.
$$

Let's return to the formula (2.5) and we will present it
in the dimensionless variables in the following form
$$
\dfrac{\sigma_{tr}}{\sigma_0}=-\dfrac{iy}{q\sqrt{\pi}}
\int\limits_{-\infty}^{\infty}\Big[1-\dfrac{q}{x}\tau+
\dfrac{e^{-q^2/4}}{2x}\big(e^{q\tau}-e^{-q\tau}\big)\Big]
\dfrac{e^{-\tau^2}d\tau}{\tau-z/q},
\eqno{(2.10)}
$$
or
$$
\dfrac{\sigma_{tr}}{\sigma_0}=-i\dfrac{y}{q}t\Big(\dfrac{z}{q}\Big)+
i\dfrac{y}{x}\lambda_C\Big(\dfrac{z}{q}\Big)-
\dfrac{iye^{-q^2/4}}{2qx\sqrt{\pi}}\int\limits_{-\infty}^{\infty}
\dfrac{e^{qt}-e^{-qt}}{t-z/q}e^{-t^2}dt.
$$

Besides, the formula (2.10) can be written down in the form
$$
\dfrac{\sigma_{tr}}{\sigma_0}=-i\dfrac{y}{q}t\Big(\dfrac{z}{q}\Big)+
i\dfrac{y}{x}\lambda_C\Big(\dfrac{z}{q}\Big)+\dfrac{iy}{2x}T(z,q),
$$
where
$$
T(z,q)=\dfrac{1}{\sqrt{\pi}}\int\limits_{-\infty}^{\infty}
\dfrac{e^{-t^2}dt}{(t-z/q)^2-q^2/4}.
$$

\begin{center}
  \bf 4. Comparison with Lindhard's formula
\end{center}

Let's consider the Lindhard's formula  (5.3.4) from \cite {Dressel} for
transverse conductivity. After limiting transition at $\eta\to 0$
from (5.3.4) we receive
$$
\hat{\sigma}^{\rm Lind}(\mathbf{q},\omega)=\dfrac{iNe^2}{\omega m}-
\dfrac{ie^2\hbar^2}{\Omega\omega m^2}\sum\limits_{\mathbf{k}}\Bigg[
\dfrac{f^0(\E_k)}{\E_{\mathbf{k+q}}-\E_{\mathbf{k}}-
\hbar(\omega+i\nu)}-
$$
$$
\hspace{5cm}-
\dfrac{f^0(\E_{k})}{\E_{\mathbf{k}}-\E_{\mathbf{k-q}}-
\hbar(\omega+i\nu)}\Bigg]|\langle\mathbf{k}+\mathbf{q}|\mathbf{p}|
\mathbf{k}\rangle_*|^2.
\eqno{(3.1)}
$$

Here $f^0(\E_k)=\dfrac {N}{v_T^3\pi^{3/2}}e^{-P^2}$ is
absolute Maxwell --- Boltzmann distribution,
$\E_{\mathbf{k}}=\dfrac{\hbar^2\mathbf{k}^2}{2m}$;
besides, the sum from (3.1) is understood as integral
$$
\dfrac{1}{\Omega}\sum\limits_{\mathbf{k}}=%\dfrac{1}{\pi^{3/2}}
\int\dfrac{\hbar^3}{m^3} d\mathbf{k}=
\int\dfrac{\hbar^3}{m^3}\dfrac{p_T^3\,d^3P}{\hbar^3}=
\int v_T^3\,d^3P.
$$

Let's present the formula (3.1) in the form
$$
\hat{\sigma}^{\rm Lind}(\mathbf{q},\omega)=\dfrac{i\sigma_0}{\omega\tau}+
\hat{\sigma}_2,
\eqno{(3.2)}
$$
where
$$
\hat{\sigma}_2=-
\dfrac{ie^2\hbar^2}{\Omega\omega m^2}\sum\limits_{\mathbf{k}}\Bigg[
\dfrac{f^0(\E_k)}{\E_{\mathbf{k+q}}-\E_{\mathbf{k}}-
\hbar(\omega+i\nu)}-
$$
$$
\hspace{5cm}-
\dfrac{f^0(\E_{k})}{\E_{\mathbf{k}}-\E_{\mathbf{k-q}}-
\hbar(\omega+i\nu)}\Bigg]|\langle\mathbf{k}+\mathbf{q}|\mathbf{p}|
\mathbf{k}\rangle_*|^2.
\eqno{(3.3)}
$$

Let us notice that
$$
\E_{\mathbf{k+q}}-\E_{\mathbf{k}}=\dfrac{\hbar^2}{2m}
\Big[q^2+2\mathbf{kq}-\dfrac{2m}{\hbar}(\omega+i \nu)\Big],
$$
$$
\E_{\mathbf{k}}-\E_{\mathbf{k-q}}=-\dfrac{\hbar^2}{2m}
\Big[q^2-2\mathbf{kq}+\dfrac{2m}{\hbar}(\omega+i \nu)\Big].
$$

Let's present the formula (3.3) in the integrated form
$$
\hat{\sigma}_2=-\dfrac{2ie^2N}{m\omega \pi^{3/2}}\int
\Big[k^2-\Big(\dfrac{\mathbf{k}\mathbf{q}}{q}\Big)^2\Big]
\times $$$$ \times
\Bigg[\dfrac{1}{q^2+2\mathbf{k}\mathbf{q}-\dfrac{2m}{\hbar}(\omega+
i/\tau)}+\dfrac{1}{q^2-2\mathbf{k}\mathbf{q}+\dfrac{2m}{\hbar}(\omega+
i/\tau)}\Bigg]e^{-P^2}d^3P.
\eqno{(3.4)}
$$

Let's transform the formula (3.4). We will direct a wave vector
$\mathbf{q}$ lengthways
$x$--component of momentum, i.e. we take $\mathbf{q}=\{k,0,0\}$, and
instead of vector $\mathbf{k}$ we will enter a dimensionless vector
$\mathbf{P}$ by the following equality
$$
\mathbf{k}=\dfrac{\mathbf{p}}{\hbar}=\dfrac{p_F}{\hbar}\mathbf{P},
\qquad p_F=mv_F.
$$

Then the first square bracket is equal to
$$
k^2-\Big(\dfrac{\mathbf{k}\mathbf{q}}{q}\Big)^2=
\dfrac{p_F^2}{\hbar^2}\Big(P^2-P_x^2\Big)=
\dfrac{p_F^2}{\hbar^2}\Big(P_y^2+P_z^2\Big)=\dfrac{p_F^2}{\hbar^2}
P_\perp^2.
$$

The second square bracket is equal to
$$
\Bigg[\dfrac{1}{q^2+2\mathbf{k}\mathbf{q}-\dfrac{2m}{\hbar}(\omega+
i/\tau)}+\dfrac{1}{q^2-2\mathbf{k}\mathbf{q}+\dfrac{2m}{\hbar}(\omega+
i/\tau)}\Bigg]=
$$
$$
=\dfrac{i\hbar}{2m\nu}\Bigg[\dfrac{1}{1-i\omega\tau+il\mathbf{kP}+
i\hbar\dfrac{k^2\tau}{2m}}-\dfrac{1}{1-i\omega\tau+il\mathbf{kP}-
i\hbar\dfrac{k^2\tau}{2m}}\Bigg]=
$$
$$
=\dfrac{\hbar^2k^2\tau}{2m^2\tau}\dfrac{1}{(1-i\omega\tau+
il\mathbf{kP})^2+\Big(\dfrac{\hbar k^2\tau}{2m}\Big)^2}.
$$

Here
$$
\dfrac{\hbar k^2\tau}{2m}=\dfrac{k^2v_T\tau}{2mv_T/\hbar}=
\dfrac{k^2l}{2k_T}=\dfrac{lkq}{2}.
$$

Now we represent integral Lindhard's term  in the form
$$
\hat{\sigma}_2=-i\sigma_0\dfrac{(lk)^2}{\omega\tau}\dfrac{1}{\pi^{3/2}}
\int \dfrac{e^{-P^2}P_\perp^2\,d^3P}{(1-i\omega\tau+il\mathbf{kP})^2
+(l\mathbf{kq}/2)^2}.
\eqno{(5.5)}
$$

The formula (3.5) for $\hat{\sigma}_2$ exactly coincides
with the formula (1.21) for $\sigma_2$. It means, that deduced in
the present work the formula for calculation of the electric
conductivity in quantum collisional plasma, does not coincide with
the corresponding formula deduced by Lindhard.

Let's write down their difference (in dimensionless parametres)
$$
\sigma_{tr}-\sigma_{tr}^{\rm Lind}=-i\sigma_0\dfrac{y}{x}\Big[1+
\dfrac{x}{q}t\Big(\dfrac{z}{q}\Big)-\lambda_c
\Big(\dfrac{z}{q}\Big)\Big]=
$$
$$
=-\sigma_0 \dfrac{y^2}{xq}t\Big(\dfrac{x+iy}{q}\Big).
$$

From this formula follows, that at $q\to\infty $ the difference
$\sigma_{tr}-\sigma_{tr}^{\rm Lind}\to 0$.

On Figs. 1 -- 10 we will give the graphic analysis of electric
conductivity. On
Figs. 1, 3 and 5 plots of the real  part of
relation $\sigma_{tr}/\sigma_0$ are presented, and on Figs. 2, 4 and 6
plots of the imaginary part of this relation  are presented as function
od dimensionless wave number $q$.

The analysis of plots on Figs. 1 - 6 shows, that at small values of
dimensionless wave number the curves corresponding to the quantum
plasma (these are curves of $1$), coincide with the curves
corresponding to classical plasma (these are curves of $3$).
Intermediate values of
dimensionless wave number, where curves of $1$ and $3$ not
coincide, make an interval of $0.1 <q <1$ in a case $x=0.1,
y=0.01$.

On Figs. 7 -- 10 dependences for real (Figs. 7 and 9) and
imaginary (Figs. 8 and 10) parts of relation $\sigma_{tr}/\sigma_0$
are presented as functions of dimensionless frequency of a field $x$ at
various values of dimensionless wave number $q$.

At
great values $x$ the curves corresponding to various values
$q$, coincide. So, in the case $y=0.01$ and small values of
dimensionless wave number $q=0.1, 0.2, 0.3$
the real and imaginary parts of
relations $\sigma_{tr}/\sigma_0$ coincide at $x> 1$. In a case $y=0.01$
and great values of dimensionless wave number
$q=2,3,4$ the real parts of relation $\sigma_{tr}/\sigma_0$,
coincide at $x> 20$, and imaginary coincide at $x> 10$.

\begin{center}\bf
  5. Conclusions
\end{center}

In the present work formulas for calculation of transverse
conductivity and permeability in quantum collisional Maxwellian
plasma  are obtained. The kinetic Wigner -- Vlasov -- Boltzmann
equation with collision integral  in
the  BGK form in coordinate space is used.

Expand of transverse  conductivity by degrees dimensionless of wave vector
is derived.

When Planck's constant
$\hbar\to 0$, the deduced formula for conductivity pass in
the formula of conductivity for classical plasma.

Comparison of the deduced formula for the transverse conductivity
with Lindhard conductivity is carried out.

The graphic analysis of the dependences of  real and imaginary
parts of transverse conductivity  on the dimensionless
wave number, and on dimensionless frequency
electric field is given.

At small values of the dimensionless wave
numbers the transverse conductivity is well described by the formula
for
classical conductivity, and at great values by Lindhard's formula.

\begin{figure}[h]  %\centering
\includegraphics[width=18.0cm, height=9.5cm]{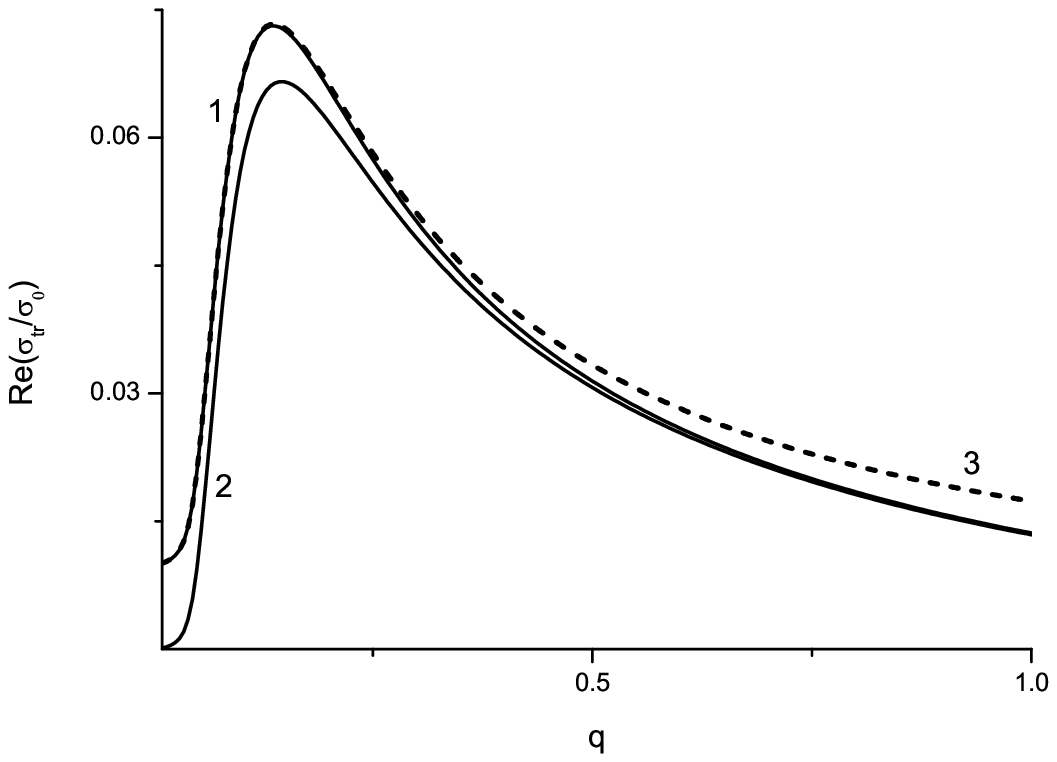}
\caption{The case: $x=0.1,y=0.01.$
Dependence  $\Re(\sigma_{tr}/\sigma_0)$ on the dimensionless wave
number $q$.}
%\end{figure}
%\begin{figure}[h]  %\centering
\includegraphics[width=18.0cm, height=9.5cm]{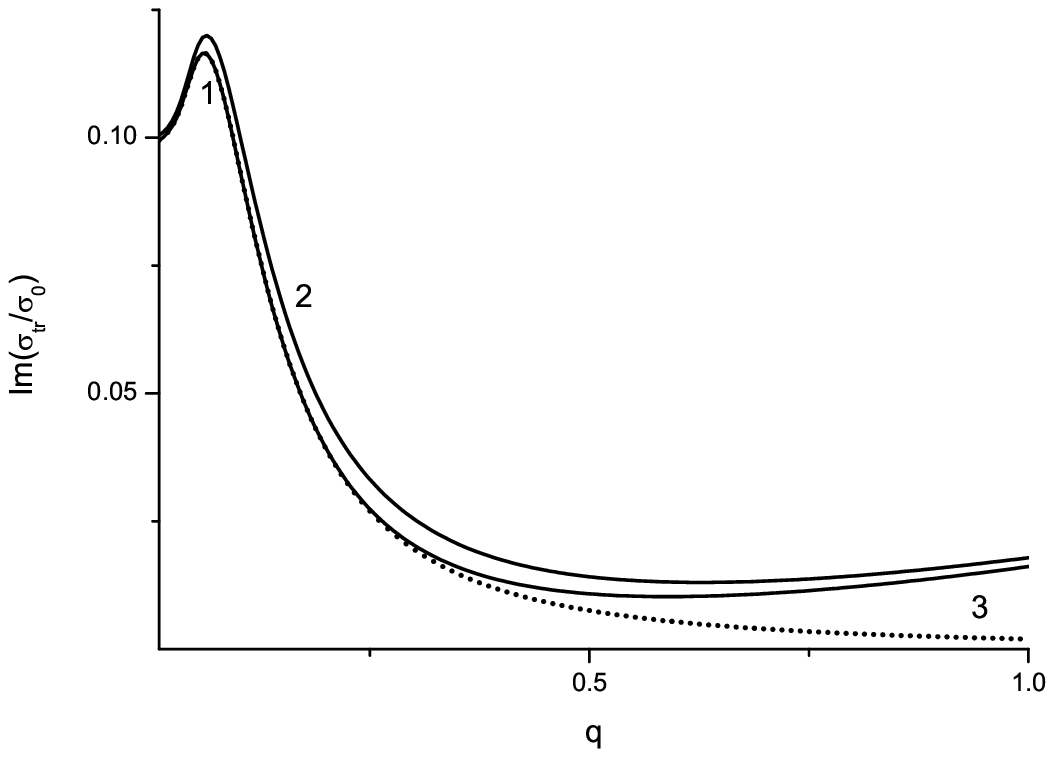}
\caption{The case: $x=0.1,y=0.01.$
Dependence  $\Im(\sigma_{tr}/\sigma_0)$ on the dimensionless wave
number $q$.}
\end{figure}

\begin{figure}[h]  %\centering
\includegraphics[width=18.0cm, height=9.5cm]{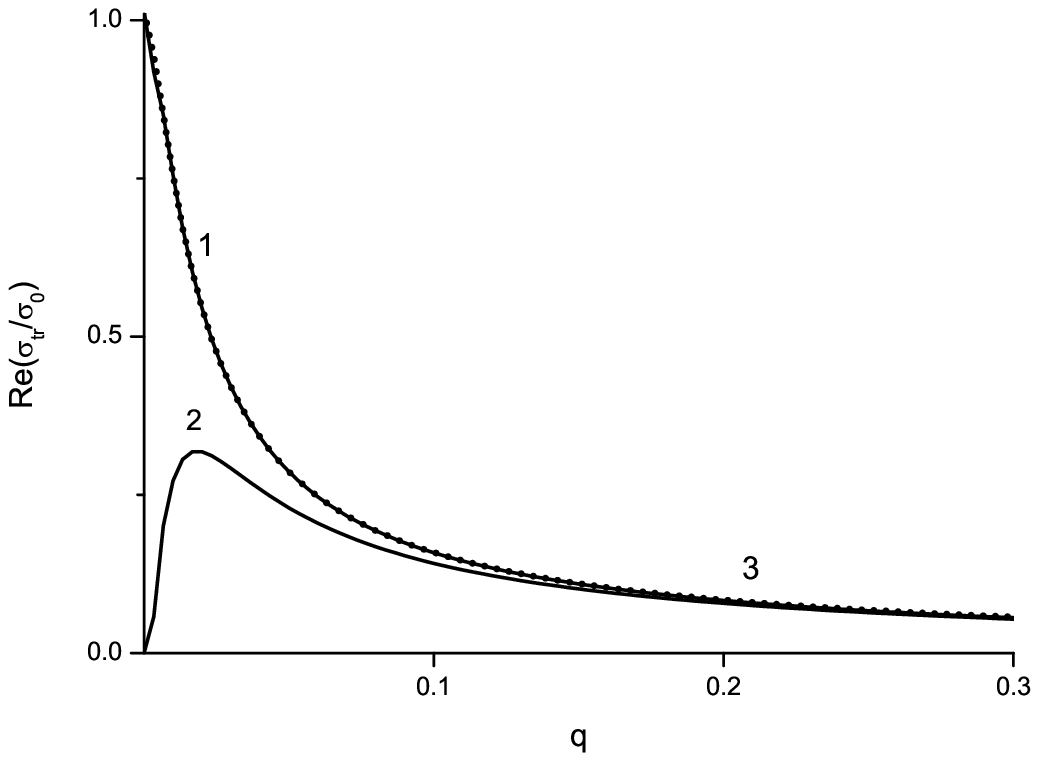}
\caption{The case: $x=0.001,y=0.01.$
Dependence  $\Re(\sigma_{tr}/\sigma_0)$ on the dimensionless wave
number $q$.}
\end{figure}
\begin{figure}[h]  %\centering
\includegraphics[width=18.0cm, height=9.5cm]{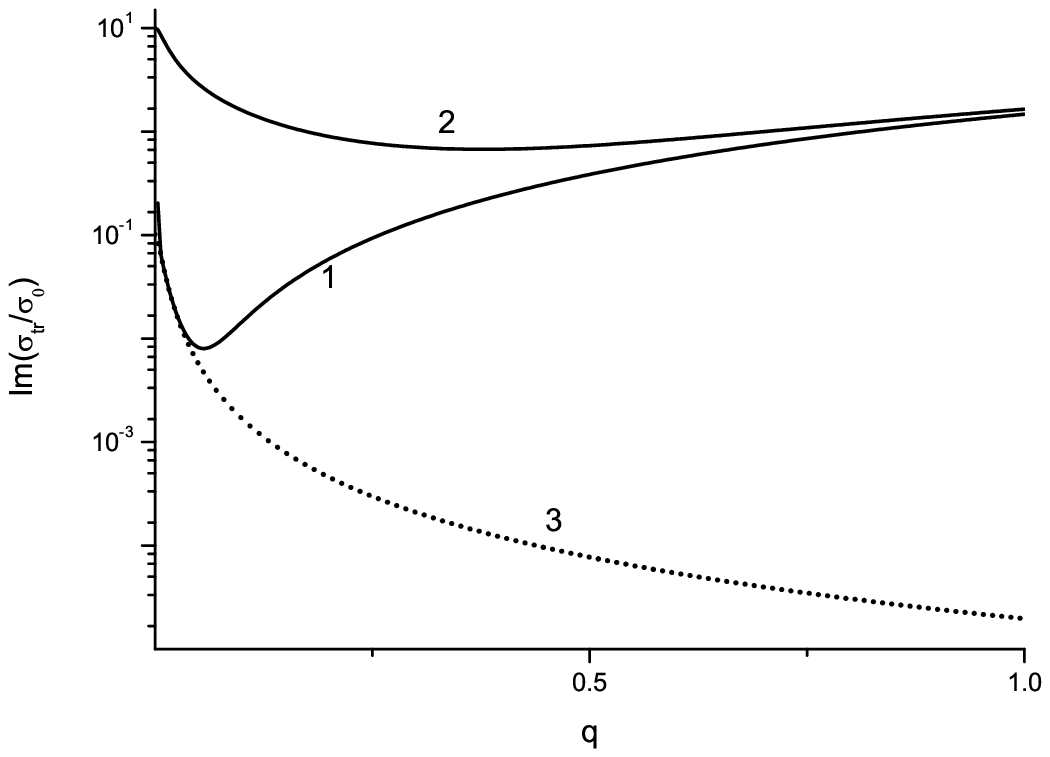}
\caption{The case: $x=0.001,y=0.01.$
Dependence  $\Im(\sigma_{tr}/\sigma_0)$ on the dimensionless wave
number $q$. }
\end{figure}

\begin{figure}[h]  %\centering
\includegraphics[width=18.0cm, height=9.5cm]{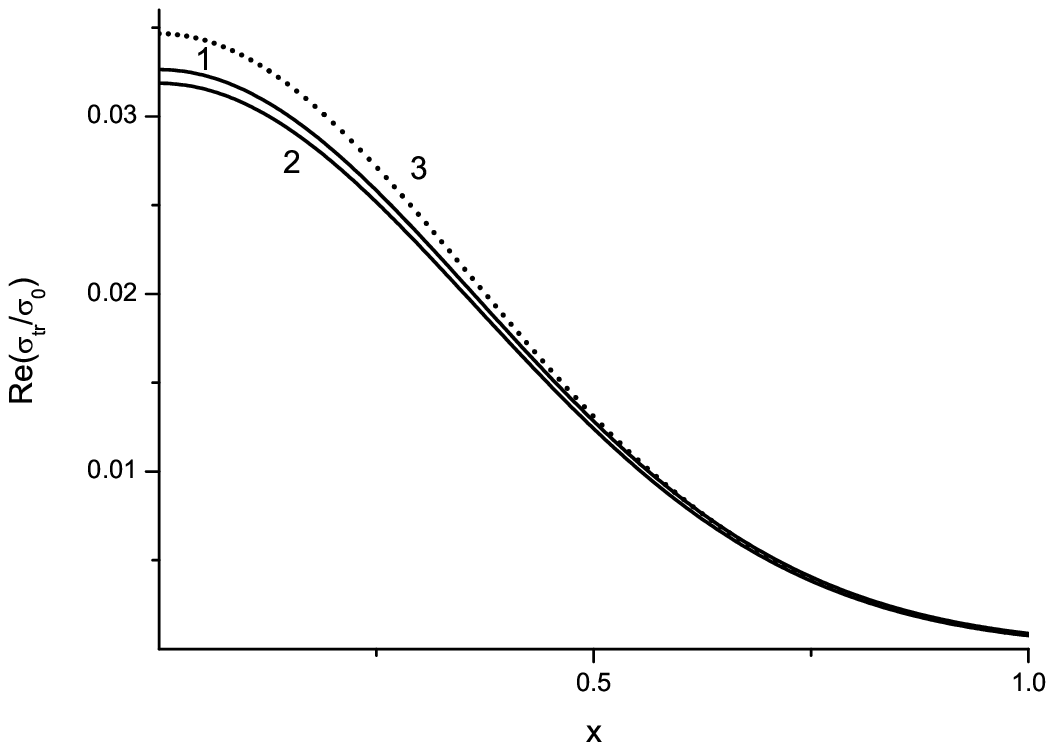}
\caption{The case: $q=0.5,y=0.01.$
Dependence  $\Re(\sigma_{tr}/\sigma_0)$ on the dimensionless frequency
$x$.}
\end{figure}

\begin{figure}[h]  %\centering
\includegraphics[width=18.0cm, height=9.5cm]{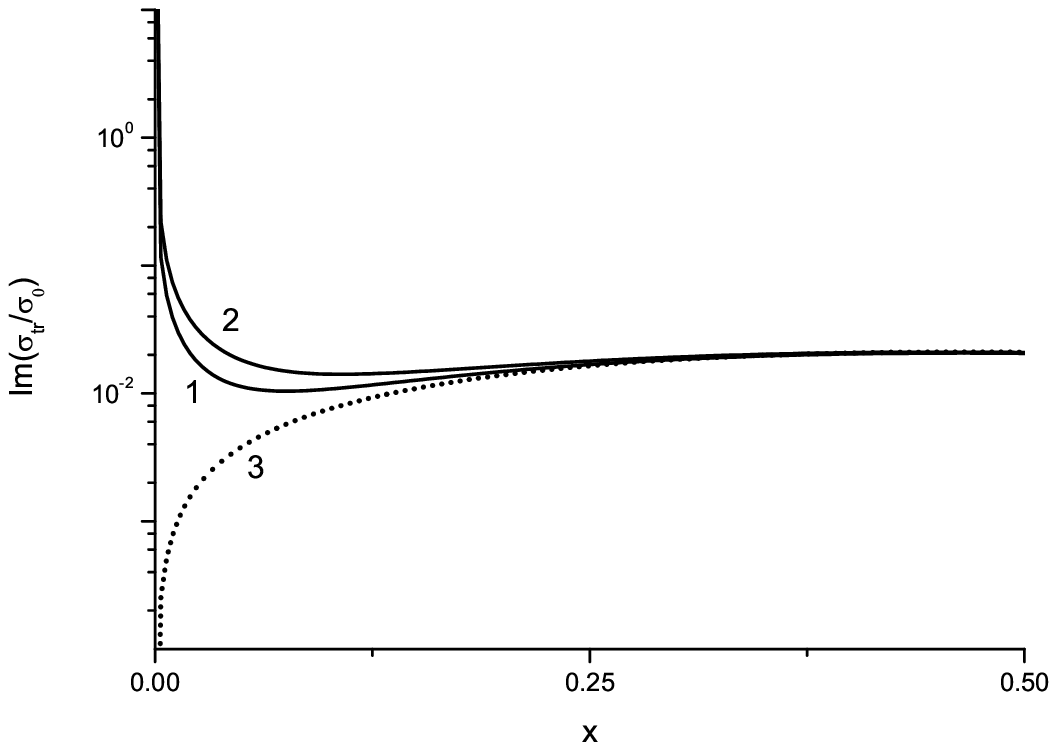}
\caption{The case: $q=0.5,y=0.01.$
Dependence  $\Im(\sigma_{tr}/\sigma_0)$ on the dimensionless frequency
$x$.}
\end{figure}

\begin{figure}[h]  %\centering
\includegraphics[width=18.0cm, height=9.5cm]{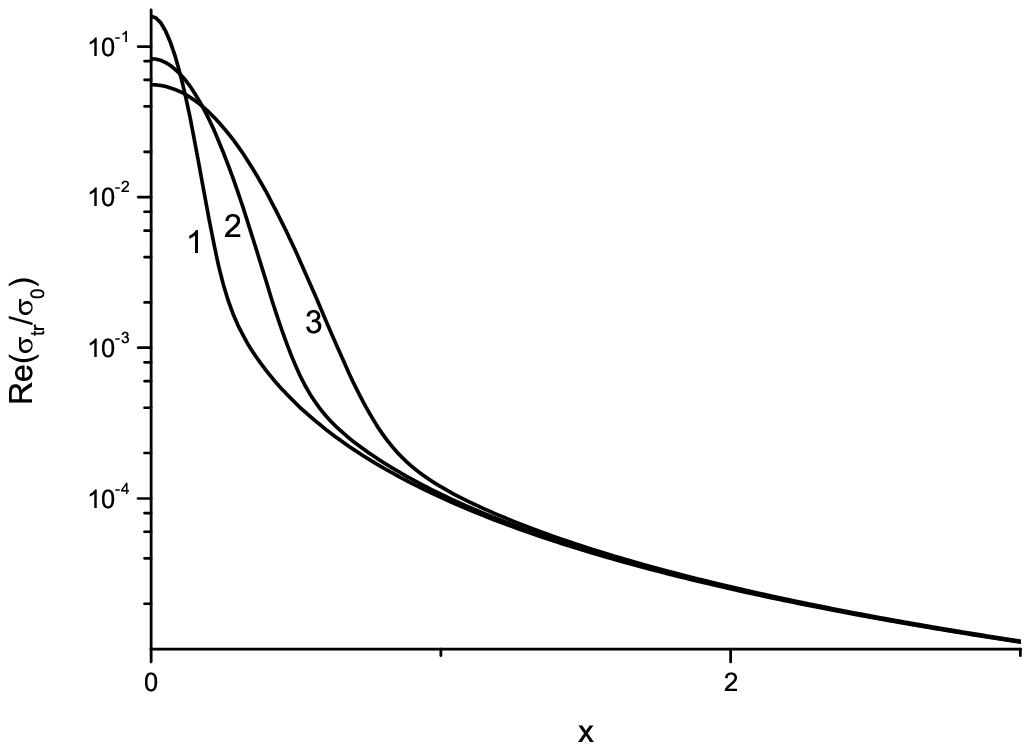}
\caption{The case: $y=0.01.$
Dependence  $\Re(\sigma_{tr}/\sigma_0)$ on the dimensionless frequency
$x$. Curves $1,2,3$ correspond to $q=0.1,0.2,0.3$.}
\end{figure}

\begin{figure}[h]  %\centering
\includegraphics[width=18.0cm, height=9.5cm]{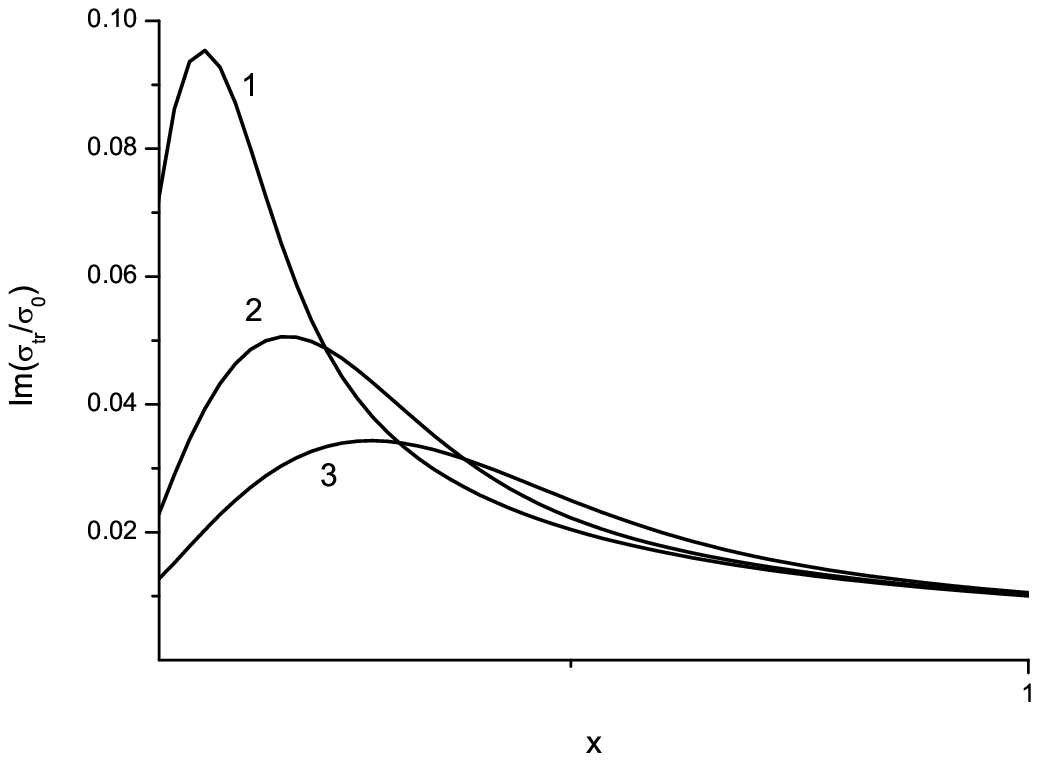}
\caption{The case: $y=0.01.$
Dependence  $\Im(\sigma_{tr}/\sigma_0)$ on the dimensionless frequency
$x$. Curves $1,2,3$ correspond to $q=0.1,0.2,0.3$.}
\end{figure}

\begin{figure}[h]  %\centering
\includegraphics[width=18.0cm, height=9.5cm]{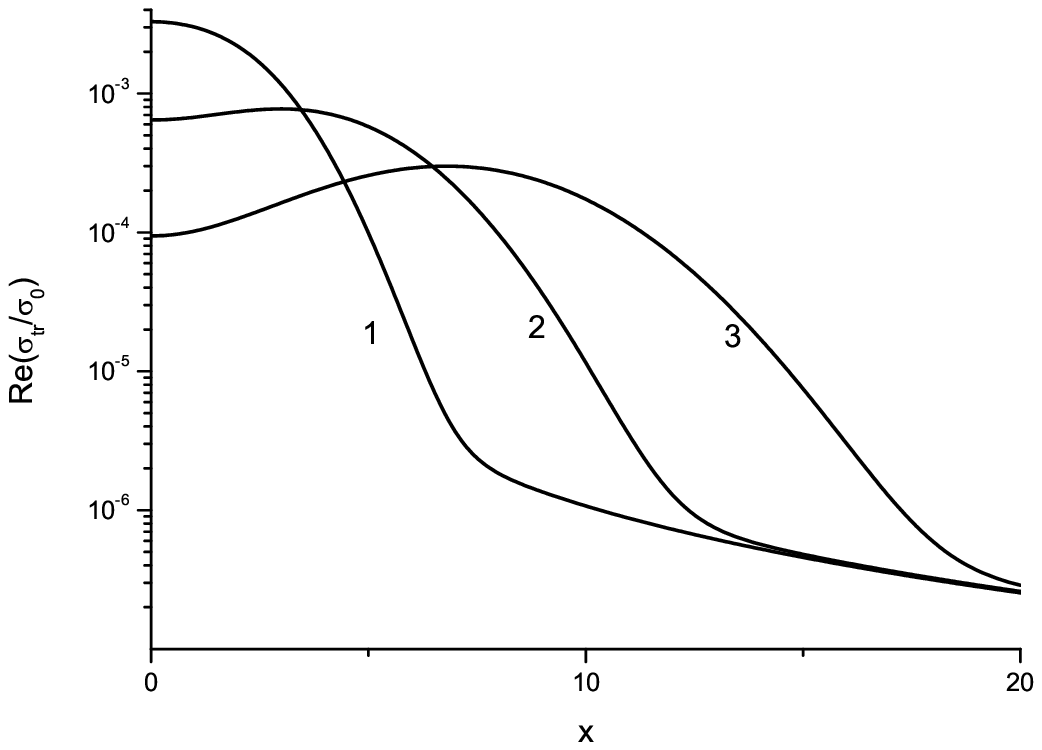}
\caption{The case: $y=0.01.$
Dependence  $\Re(\sigma_{tr}/\sigma_0)$ on the dimensionless frequency
$x$. Curves $1,2,3$ correspond to $q=2,3,4$.}
\end{figure}

\begin{figure}[h]  %\centering
\includegraphics[width=18.0cm, height=9.5cm]{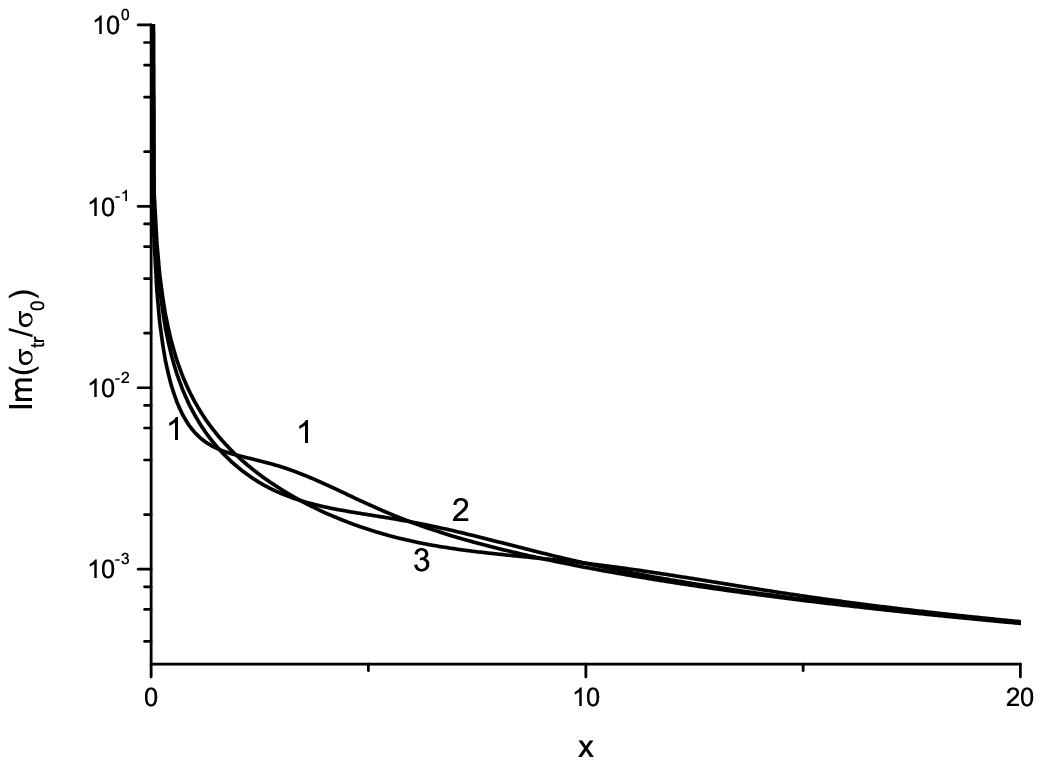}
\caption{The case: $y=0.01.$
Dependence  $\Im(\sigma_{tr}/\sigma_0)$ on the dimensionless frequency
$x$. Curves $1,2,3$ correspond to $q=2,3,4$.}
\end{figure}

\end{document}